\documentclass{aa}
\usepackage{psfig}
\psfull

\def\gsim{ \lower .75ex \hbox{$\sim$} \llap{\raise .27ex \hbox{$>$}} } 
\def\lsim{ \lower .75ex\hbox{$\sim$} \llap{\raise .27ex \hbox{$<$}} } 

\authorrunning{S. Campana et al.}
\titlerunning{80 d periodicity in XTE J1946+274}

\begin{document} 

\title{Evidence for an $\sim 80$ day periodicity in the X--ray transient 
pulsar XTE J1946+274}

\author{Sergio Campana\inst{1} \and Gianluca Israel\inst{2} \and Luigi
Stella\inst{2}}

\institute{Osservatorio Astronomico di Brera, Via Bianchi 46, I--23807
Merate (Lc), Italy
\and
Osservatorio Astronomico di Roma, via Frascati 33,
I--00040 Monteporzio Catone, Roma, Italy
}

\offprints{S. Campana}
\mail{campana@merate.mi.astro.it}

\date{Received ; Accepted }

\maketitle

\begin{abstract}
We report evidence for an $\sim 80$ d periodicity in the X--ray
flux of the hard transient pulsar XTE J1946+274. The 1.3--12 
keV light curve obtained with the RossiXTE All Sky Monitor shows five 
regularly spaced flares over a $\sim 1$~year baseline starting from the
outburst onset in Sept. 1998. The first and strongest flare is
somewhat longer than the subsequent four flares, which recur in a fairly 
periodic fashion. This suggests that the profile of the first flare is 
dominated by the time variability of the Be star ejection episode, while the
following four flares are primarily caused by the neutron star motion
along an eccentric orbit. 
\end{abstract}

\section{Introduction}

The hard X--ray transient (HXRT) source XTE J1946+274 was discovered with the
RossiXTE All Sky Monitor (ASM) during a scan of the Vul-Cyg region on
Sept. 5, 1998. The source was detected at a flux level of $\sim 13$ mCrab
(2--12 keV; Smith \& Takeshima 1998), which raised steeply in the
following days, reaching a peak of $\sim 110$ mCrab around Sept. 17, 1998
(Takeshima \& Chakrabarty 1998). X--ray pulsations at 15.8 s were
discovered by BATSE (GRO J1944+26, Wilson et al. 1998a) and subsequently 
confirmed through RossiXTE pointed observations (Smith \& Takeshima 1998). The
1998 outburst of XTE J1946+274 was extensively monitored also through a
BeppoSAX observational campaign (Campana et al. 2000; Santangelo et al.
2000).

The position error circle obtained with RossiXTE ($2'.4$ radius, 90\%
confidence level; Takeshima \& Chakra\-barty 1998) was reduced to $30''$
(95\% confidence level) through pointed observations with BeppoSAX Narrow
Field Instruments (Campana et al. 1998). The best X--ray position is
R.A.= 19$^{\rm h}$ 45$^{\rm m}$ 38$^{\rm s}$ and Dec.= +27$^{\rm o}$
$21'.5$ (equinox 2000). XTE J1946+274 lies in the error box of the 1976
Ariel V transient 3A 1942+274 (Warwick et al. 1981).  
Assuming $\sim 1000$ HXRTs in the Galaxy (e.g. Bildsten et al. 1997), 
we estimate a chance probability of $\sim 7\%$ of 
of finding a new transient within the 3A 1942+274 error box.
This probability is such that we cannot infer a firm association
of the two sources.

The likely optical counterpart has been recently identified (Verrecchia 
et al. 2000) with a ${\rm R} \sim 14$ mag Be star, showing a strong 
$H_{\alpha}$ emission line. 
Earlier reports of a B counterpart (Israel, Polcaro \& 
Covino 1998; Ghavamian \& Garcia 1998) has been ruled out by the BeppoSAX
error circle, lying $10''$ outside. 

All these characteristics testify that XTE J1946+274 is a Be star HXRT. 
Here we report the discovery of an $\sim 80$~d
X--ray flux modulation in RossiXTE-ASM data of the source and argue that
this (or its double) likely represents the orbital period of the system.

\begin{figure*}[!htb]
\centerline{\psfig{figure=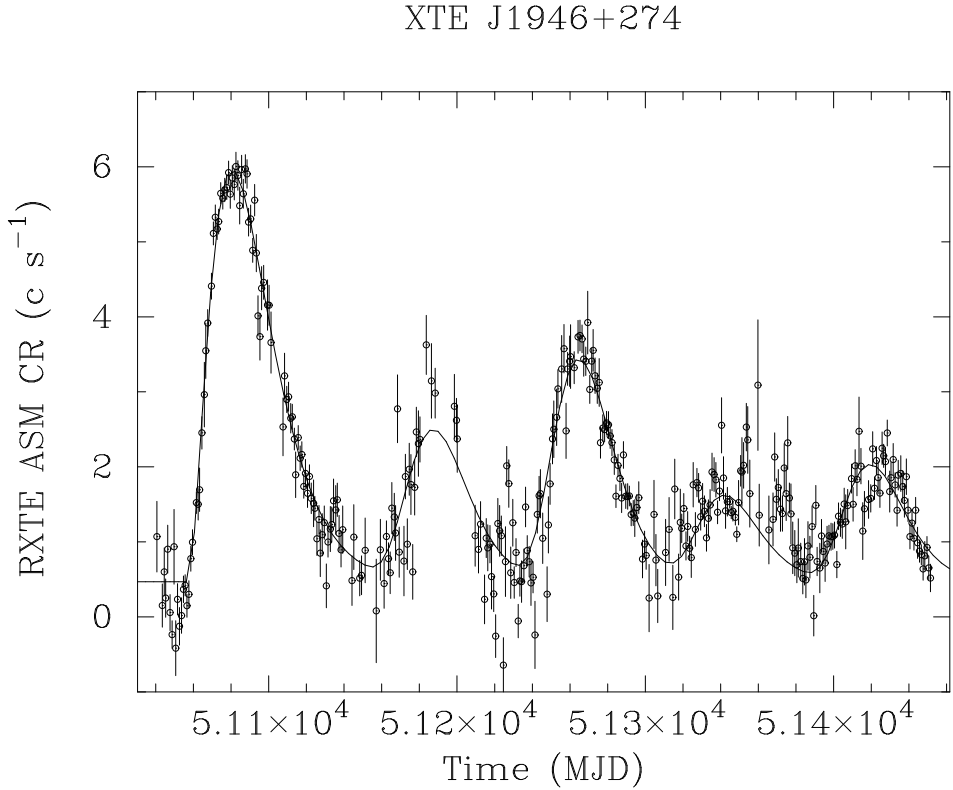,angle=-90,width=18.0cm}}
\caption{Light curve of the XTE J1946+274 as observed with the
RossiXTE-ASM. The continuous line shows the fit described in the text. 
Points with an error larger than 0.5 c s$^{-1}$ have been discarded.}
\end{figure*}

\section{Period determination}

The ASM (Levine et al. 1996) on board the RossiXTE (Bradt, Rothschild \&
Swank 1993) routinely scans about 80\% of the X--ray sky every orbit. It
consists of three Scanning Shadow Cameras with a 1.3--12 keV energy band,
an intrinsic angular resolution of a few arcmin and a large field of view
($6^{\rm o}\times 90^{\rm o}$ FWHM). A sensitivity of $\sim 5-10$ mCrab is
reached over one day. The intensity is calculated in three energy bands
(1.3--3, 3--5 and 5--12 keV) and normalized in units of source count
at the center of the field of view. Errors are computed 
considering the uncertainties due to the counting statistics and a 
$\sim 2\%$ systematic error obtained from the Crab calibration.
A description of the ASM and its light curves can be found in Levine et
al. (1996) and Levine (1998)\footnote{See the web page
http://heasarc.gsfc.nasa.gov/docs/xte/\-asm\_products.html\#access}.

Fig. 1 shows the light curve of the XTE J1946+274 outburst, which
began in Sept. 1998. The source took $\sim 25$ d to reach the peak and
then decayed smoothly, leading to a markedly asymmetric profile during
the first $\sim 80$~d of the outburst. Following the main flare four
secondary flares are evident in Fig. 1.
Fig. 2 shows the light curves in the three ASM energy channels.  The
low energy curve contains only a small signal, a result of the fact that
the source is heavily absorbed. The hardness ratio from the higher
energy channels is consistent with being constant.

\begin{figure*}[!htb]
\centerline{\psfig{figure=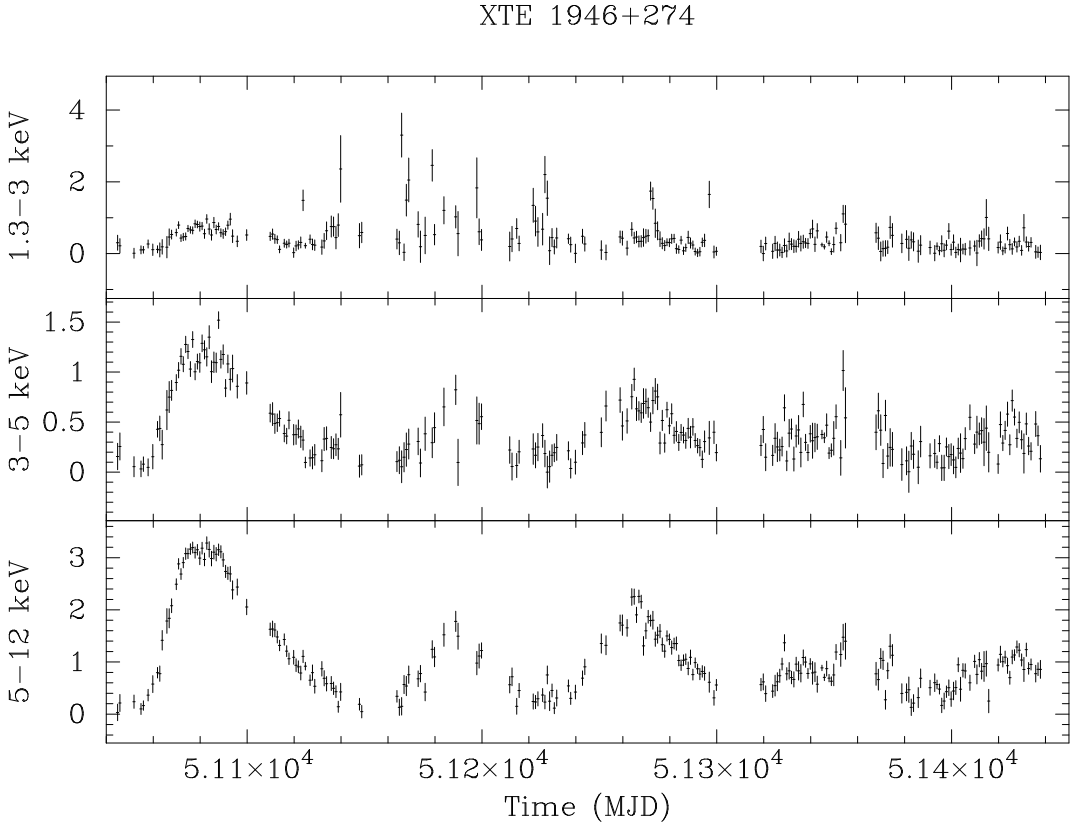,angle=-90,width=16.0cm}}
\caption{Light curve of the XTE J1946+274 in the three energy bands of the
RossiXTE-ASM.}
\end{figure*}

The overall behaviour of the light curve is suggestive of a relatively
intense outburst in the initial phases, during which the neutron star mass
capture rate was likely driven primarily by the time variations in the mass
ejection from the Be star. 
The recurrent behaviour and lower flux of the following 
four flares suggest instead that later in the outburst the mass capture
rate variations were induced mainly by the motion of the neutron star in
an eccentric orbit. 
The fairly regular $\sim 80$ d recurrence of the four secondary 
flares is more apparent by looking at the light curve minima, which are
evidently less influenced by the varying amplitude and shape of the
outburst.
 
A simple power spectrum analysis of the four secondary flares 
(from MJD 51140 to MJD 51450) reveals a significant
periodicity ($\sim 3.3\,\sigma$ in the 10--250 d range) at $\sim 
73\pm 5$ d.
The inclusion of the first flare (from MJD 51040 to MJD 51450)
lowers considerably the significance.
Similarly, by adopting an epoch folding analysis of the light curve 
we detect a significant power at a frequency corresponding to 
$\sim 79\pm 6$ d. The inclusion of the first flare causes the folding 
peak to shift to $\sim 87\pm 6$ d.
We also fitted the light curve of the last four flares 
with a sinusoid plus a constant: we determined a best fit period
of $\sim 77$ d. The same fit on the total light curve provides 
instead a period of $\sim 95$ d.

The discrepancy between the periods determined with and without the first
flare arises because this starts some $\sim 30$~d before the
extrapolation of the ephemerides derived from the subsequent four flares. 
In order to model the flare profiles more accurately we adopted a model
consisting of a smooth burst profile with a power law rise and an
exponential decay, namely $CR=CR_0\,\Bigl((t-t_0)/(t_{\rm
c}\,\alpha)\Bigr)^{\alpha}\, \exp{(t-t_0)/t_{\rm c}}$. Here $t_0$ is the
starting time of the flare, $CR_0$ is the count rate normalization, 
$\alpha$ the power law
slope of the rise and $t_{\rm c}$ the exponential decay timescale; 
for time smaller than $t_0$ the function value is 0. This
model was fitted to each of the four secondary flares by forcing all
parameters but the normalisation to be the same and
by imposing a periodic recurrence of the outbursts, with the period a
free parameter. Only the first flare, for the reasons described above, was
allowed a different rise and decay profile and a shift in time.

Fig. 1 shows the best fit obtained in this way. The reduced $\chi^2$ is 
2.5. The rising exponents are 2.0 and 8.2 and the decay times $\alpha=12$ 
and 6 d for the first and the other flares, respectively. These values are
significantly different and confirm `a posteriori' the differences 
between the first and the subsequent flares.
The best separation between the flares is 77.3 d with a nominal 
statistical uncertainty of $\sim 1$ d (90\% confidence level). 
We speculate that the slightly different result obtained with epoch folding
search and the power spectrum density are due to the small number of points,
the strength of the first and third flares and the width of the 
second and forth flares for which the maximum cannot be identified with 
certainty.
The onset of the first flare is shifted by --7 d with respect to the 
other flares, moreover the different behaviour of its decay ma\-kes it occur 
some 30 d before the value extrapolated by the following flares.
The approximate epoch of the maxima are MJD 51080.5, 51187.4, 51264.8, 
51342.1 and 51419.5. 

\section{Discussion}

In recent years, growing evidence has been gathered that the two classes of 
outburst from HXRT sources (type I: normal; type II: giant,
cf. Stella, White \& Rosner 1986) are associated with wind and disk 
accretion, respectively (Bildsten et al. 1997). 
Raguzova \& Lipunov (1998) described the Be disk-fed  
accretion in HXRTs, finding that the outburst peaks occur at phase 0--0.5, 
depending on the wind characteristics and/or orbital eccentricity 
(i.e. the wind rose effect). 
The fairly regular spacing of the last four flares in XTE J1946+26
hints for their classification as Type I outbursts, even if the flux at 
minimum does not drop to very low values as, e.g., in the case of 
V 0332+53 (Stella, White \& Rosner 1986).
The different rise and decay timescales, as well as the larger 
intensity, make however the first flare markedly different from the others.
This would argue in favour of a type II outburst.
Despite the small number of cases, type II outburst peaks have been observed 
to be delayed in orbital phase with respect to periastron
(e.g. 4U 0115+63, Whitlock et al. 1989; A 0535+26, Motch et al. 1991; 
Bildsten et al. 1997, see however 2S 1417--624 ibidem) and sometimes they 
last for several orbital cycles (e.g. V0332+53; Stella, White \& 
Rosner 1986).
The phasing of this outburst should therefore be dictated mainly by the 
time variability of the Be star mass outflow rate, especially in the 
first phases of the shell ejection episode.

As can be noted in Fig. 1, the first, third and fifth flares are 
stronger than the second and the forth. This might suggest that the 
neutron star is orbiting the Be companion in an inclined orbit and 
it crosses twice the Be disk plane, giving rise to two outbursts 
per orbit. In this case the orbital period would be $\sim 155$ d. 
Two flares per orbit
have been already observed in 4U 1907+097 (with an orbital period 
of 8.4 d) at orbital phases $\sim 0.04$ and $\sim 0.48$ (Makashima 
et al. 1984) and in GRO J2058+42 (110 d) with the two flares separated
in phase by $\sim 0.5$ (Wilson et al. 1998b). 

\section{Conclusions}

We discovered in the RossiXTE-ASM data of XTE J1946 +274 an $\sim 80$ d 
modulation. This periodicity likely represent the orbital period of the 
system or, perhaps, half this value if the neutron star is orbiting
the Be out of its shell ejection plane. 
A correlation between the spin and orbital periods of Be star/X--ray 
pulsar binary systems was described by Corbet (1984, 1986).
Despite the considerable scatter in this relationship, given the 15.8 s 
spin period of XTE J1946+274, one would obtain for an orbital period of 
$\sim 77$~d a moderate eccentricity ($e\lsim 0.4$) and for its double
a somewhat more eccentric system ($e\lsim 0.6$).
 
An orbital solution, affording an independent accurate measurement of the
orbital period and eccentricity, might be obtained from the analysis of
the pulse arrival times during the pointed RossiXTE observations.

\begin{acknowledgements}
We acknowledge the use of quick-look results provided by the RossiXTE-ASM team
and useful comments from H. Bradt and an anonymous referee.
\end{acknowledgements}

\end{document}